\documentclass[twocolumn]{aastex63}

\usepackage{amsmath}	
\usepackage{amssymb}	
\usepackage{enumitem}
\usepackage{verbatim} 
\usepackage[caption=false]{subfig}

\newcommand{\mum}{$\mu$m}

\newcommand{\bdepthoccundilTESS}{$482_{-41}^{+39}$} 

\newcommand{\bperiod}{$1.274928\pm0.000011$} 
\newcommand{\bphasecurveatmosphericTESS}{$418\pm34$} 
\newcommand{\bphasecurveatmosphericshiftTESS}{$-0.022\pm0.014$} 
\newcommand{\bphasecurveellipsoidalTESS}{$8_{-6}^{+12}$} 
 
\newcommand{\tmptdayy}{$3012\substack{+40 \\ -42}$}
\newcommand{\tmptnigh}{$2022\substack{+254 \\ -602}$}
\newcommand{\tmptcont}{$33\substack{+20 \\ -8}$}
\newcommand{\phasshft}{$6.3\substack{+3.9 \\ -3.9}$}

\received{September 6, 2019}
\accepted{January 4, 2021}
\submitjournal{AJ}

\begin{document}

\title{\textit{TESS} observations of the WASP-121\,b phase curve}
\shorttitle{\textit{TESS} observations of the WASP-121\,b phase curve}

\correspondingauthor{Tansu Daylan}
\email{tdaylan@mit.edu}
\author[0000-0002-6939-9211]{Tansu Daylan}
\affil{Department of Physics and Kavli Institute for Astrophysics and Space Research, Massachusetts Institute of Technology, Cambridge, MA 02139, USA}
\affil{Kavli Fellow}

\author[0000-0002-3164-9086]{Maximilian N.\ G{\"u}nther}
\affil{Department of Physics and Kavli Institute for Astrophysics and Space Research, Massachusetts Institute of Technology, Cambridge, MA 02139, USA}
\affil{Juan Carlos Torres Fellow}

\author[0000-0001-5442-1300]{Thomas Mikal-Evans}
\affil{Department of Physics and Kavli Institute for Astrophysics and Space Research, Massachusetts Institute of Technology, Cambridge, MA 02139, USA}

\author[0000-0001-6050-7645]{David K. Sing}
\affil{Department of Earth and Planetary Sciences and Department of Physics \& Astronomy, Johns Hopkins University, Baltimore, MD, USA}

\author[0000-0001-9665-8429]{Ian Wong}
\affil{Department of Earth, Atmospheric, and Planetary Sciences, Massachusetts Institute of Technology, Cambridge, MA, 02139, USA}
\affil{51 Pegasi b Fellow}

\author[0000-0002-1836-3120]{Avi Shporer}
\affil{Department of Physics and Kavli Institute for Astrophysics and Space Research, Massachusetts Institute of Technology, Cambridge, MA 02139, USA}

\author[0000-0002-8052-3893]{Prajwal Niraula}
\affil{Department of Earth, Atmospheric, and Planetary Sciences, Massachusetts Institute of Technology, Cambridge, MA, 02139, USA}

\author{Julien de Wit}
\affil{Department of Earth, Atmospheric, and Planetary Sciences, Massachusetts Institute of Technology, Cambridge, MA, 02139, USA}

\author[0000-0002-9076-6901]{Daniel D. B. Koll}
\affil{Department of Earth, Atmospheric, and Planetary Sciences, Massachusetts Institute of Technology, Cambridge, MA, 02139, USA}

\author{Vivien Parmentier}
\affil{Department of Physics, Oxford University, OX1 2JD, United Kingdom}

\author{Tara Fetherolf}
\affil{Department of Earth and Planetary Sciences, University of California, Riverside, CA 92521, USA}

\author[0000-0002-7084-0529]{Stephen R.~Kane}
\affil{Department of Earth and Planetary Sciences, University of California, Riverside, CA 92521, USA}

\author[0000-0003-2058-6662]{George~R.~Ricker}
\affil{Department of Physics and Kavli Institute for Astrophysics and Space Research, Massachusetts Institute of Technology, Cambridge, MA 02139, USA}

\author[0000-0001-6763-6562]{Roland~Vanderspek}
\affil{Department of Physics and Kavli Institute for Astrophysics and Space Research, Massachusetts Institute of Technology, Cambridge, MA 02139, USA}

\author[0000-0002-6892-6948]{S.~Seager}
\affil{Department of Physics and Kavli Institute for Astrophysics and Space Research, Massachusetts Institute of Technology, Cambridge, MA 02139, USA}
\affil{Department of Earth, Atmospheric, and Planetary Sciences, Massachusetts Institute of Technology, Cambridge, MA, 02139, USA}
\affiliation{Department of Aeronautics and Astronautics, MIT, 77 Massachusetts Avenue, Cambridge, MA 02139, USA}

\author[0000-0002-4265-047X]{Joshua~N.~Winn}
\affiliation{Department of Astrophysical Sciences, Princeton University, 4 Ivy Lane, Princeton, NJ 08544, USA}

\author{Jon~M.~Jenkins}
\affiliation{NASA Ames Research Center, Moffett Field, CA, 94035, USA}

\author{Douglas~A.~Caldwell}
\affiliation{NASA Ames Research Center, Moffett Field, CA, 94035, USA}
\affiliation{SETI Institute, Mountain View, CA 94043, USA}

\author[0000-0002-9003-484X]{David~Charbonneau}
\affiliation{Center for Astrophysics | Harvard \& Smithsonian, 60 Garden St, Cambridge, MA 02138, USA}

\author{Christopher~E.~Henze}
\affiliation{NASA Ames Research Center, Moffett Field, CA, 94035, USA}

\author[0000-0001-8120-7457]{Martin~Paegert}
\affiliation{Center for Astrophysics | Harvard \& Smithsonian, 60 Garden St, Cambridge, MA 02138, USA}

\author{Stephen Rinehart}
\affiliation{NASA Goddard Space Flight Center, 8800, Greenbelt Road, MD, USA}

\author{Mark~Rose}
\affiliation{NASA Ames Research Center, Moffett Field, CA, 94035, USA}

\author{Lizhou~Sha}
\affil{Department of Physics and Kavli Institute for Astrophysics and Space Research, Massachusetts Institute of Technology, Cambridge, MA 02139, USA}

\author{Elisa Quintana}
\affiliation{NASA Goddard Space Flight Center, 8800, Greenbelt Road, MD, USA}

\author{Jesus Noel Villasenor}
\affiliation{Department of Physics and Kavli Institute for Astrophysics and Space Research, Massachusetts Institute of Technology, Cambridge, MA 02139, USA}


\begin{abstract}

We study the red-optical photometry of the ultra-hot Jupiter WASP-121\,b as observed by the Transiting Exoplanet Survey Satellite (\textit{TESS}) and model its atmosphere through a radiative transfer simulation. Given its short orbital period of $\sim1.275$ days, inflated state and bright host star, WASP-121\,b is exceptionally favorable for detailed atmospheric characterization. Towards this purpose, we use \texttt{allesfitter} to characterize its full red-optical phase curve, including the planetary phase modulation and the secondary eclipse. We measure the day and nightside brightness temperatures in the \textit{TESS} passband as \tmptdayy{} K and \tmptnigh{} K, respectively, and do not find a statistically significant phase shift between the brightest and substellar points. This is consistent with an inefficient heat recirculation on the planet. We then perform an atmospheric retrieval analysis to infer the dayside atmospheric properties of WASP-121\,b such as its bulk composition, albedo and heat recirculation. We confirm the temperature inversion in the atmosphere and suggest H$^-$, TiO and VO as potential causes of the inversion, absorbing heat at optical wavelengths at low pressures. Future \textit{HST} and \textit{JWST} observations of WASP-121\,b will benefit from its first full phase curve measured by \textit{TESS}.

\end{abstract}


\section{Introduction}

A planet's occultation as it passes behind its host star, i.e., the secondary eclipse, reveals atmospheric characteristics such as the dayside temperature and reflectivity \citep{Agol+2010, Line+2013}. Furthermore, changes in the brightness of a planet and star system as a function of orbital phase (i.e., the phase curve), contain information about the dynamics of the orbit as well as the thermal state on the nightside and heat recirculation on the planet \citep{CowanAgol2008, Angerhausen+2015, Shporer2017, Zhang+2018, Shporer+2019}.

WASP-121\,b is a transiting exoplanet with a period of \bperiod{} days discovered by \citet{Delrez+2016}. The host star WASP-121 has the \textit{TESS} Input Catalog (TIC) number 22529346 and is a F6V type star with radius of $1.52\pm0.06$~R$_\odot$, mass of $1.45\pm0.25$~M$_\odot$, and effective temperature of $6776\pm134$~K \citep{Stassun+2019}. WASP-121\,b belongs to a class of exoplanets known as ultra-hot Jupiters that orbit their host stars with short ($\sim 1$ day) periods and have dayside temperatures larger than 2200~K. Given its short orbital period, it is likely that WASP-121\,b is locked to its host star via tidal interactions. Furthermore, WASP-121\,b has various features that make it an exceptional exoplanet. First, its short orbital period and hot host star cause it to be highly irradiated, raising its dayside near-infrared brightness temperature to $\sim$2700 K \citep{Evans+2017}. Second, it has a spin-orbit angle of $258^\circ\pm{5^\circ}$ \citep{Delrez+2016}, indicating that the planet's orbit is nearly anti-aligned with the stellar spin axis. Third, given its mass and radius of $\sim$1.2 and $\sim$1.9 times that of Jupiter \citep{Delrez+2016}, respectively, WASP-121\,b is an inflated exoplanet that nearly fills its Roche lobe. This means that it should be tidally disrupted and merge with its host star in the next few hundred million years. Furthermore, given its bright host star with a V magnitude of 10.5, aforementioned inflated state and availability of previously collected spectroscopic data, WASP-121\,b is an interesting target for detailed atmospheric characterization. The James Webb Space Telescope (\textit{JWST}) is expected to further characterize the atmosphere of WASP-121\,b in the near and far-infrared via emission spectroscopy. 

Previous work on WASP-121\,b found evidence for emission and absorption features as opposed to a featureless blackbody spectrum \citep{Mikal-Evans+2020}. In \citet{Evans+2017}, 1.4 $\mu$m infrared emission from hot H$_2$O molecules was detected in the emission spectrum of WASP-121\,b using the Hubble Space Telescope (\textit{HST}) Wide Field Camera 3 (WFC3) G102 grism. Along with the Spitzer measurements, this allowed atmospheric retrieval of WASP-121\,b's dayside. That the H$_2$O feature was observed in emission rather than absorption, is a potential indication that the upper layers of the atmosphere are hotter and that the temperature profile of the atmosphere is inverted. This requires the existence of a stratosphere (i.e., the layer of the atmosphere, where the temperature increases with altitude, which is significantly heated at low pressure). This temperature inversion can be realized by metal oxides such as Vanadium Oxide (VO) or Titanium Oxide (TiO) due to their strong absorption in the optical wavelengths. These molecules are commonly found in brown dwarfs and require high temperatures to be in gaseous form, such as that available on WASP-121\,b. Moreover, an atmospheric transmission spectrum was observed using the Space Telescope Imaging Spectrograph (STIS) on \textit{HST}, where evidence for VO was found \citep{Evans+2018}. More recently, an emission spectrum for WASP-121\,b in the 0.8--1.1 $\mu$m passband was collected using \textit{HST} \citep{Mikal-Evans+2019}. This provided evidence for the contribution of H$^-$ to the optical opacity of WASP-121\,b's atmosphere. Furthermore in \citet{Sing+2019} UV transmission spectroscopy was performed on WASP-121\,b as part of the Panchromatic Comparative Exoplanet Treasury (PanCET) survey, revealing heavy ionized metals such as Mg and Fe in gas form, implying that they escape the planet instead of forming condensates.

In this work, we study the full phase curve of the WASP-121 system in the red-optical (i.e., 0.6--0.95 \mum) passband as measured by the Transiting Exoplanet Survey Satellite (\textit{TESS}; \citealt{Ricker+2014}) in order to characterize the planetary modulation (i.e., thermal and reflected emission from the planet). We then discuss the implications of these measurements regarding the atmospheric properties of WASP-121\,b, including optical reflectivity and heat recirculation.

The \textit{TESS} data on WASP-121\,b is important mainly for two reasons. First, \textit{TESS} extends the wavelength coverage of the available data on the system by 0.2 $\mu$m towards shorter wavelengths as compared to the previously collected G102 data \citep{Mikal-Evans+2019}. Interestingly, the \textit{TESS} passband contains emission bands of TiO and VO, giving us a probe of their role in the temperature inversion. Second, the \textit{TESS} data contain the full phase curve of the system, allowing inference of the nightside temperature as well as any phase shift between the bright spot (i.e., longitude of highest brightness) and the substellar point.

The rest of the paper is as follows. The data taken by \textit{TESS} are presented in Section \ref{sect:data}. Our methods of data analysis are presented in Section \ref{sect:meth}. We then present our results in Section \ref{sect:resu} and end with a discussion and conclusion in Sections \ref{sect:disc} and \ref{sect:conc}.

\section{Data}
\label{sect:data}

The WASP-121 system was observed by the \textit{TESS} mission at 2 minute cadence during Sector 7. The observations were carried out with camera 3 and the charge coupled device (CCD) 2, between 7 January 2019 and 2 February 2019, spanning orbits 21 and 22.

These short cadence target pixel files were analyzed by the Science Processing Operations Center (SPOC) pipeline \citep{Jenkins+2010, Jenkins+2016, Smith+2012, Stumpe+2014} to produce simple aperture photometry (SAP) and presearch data conditioning (PDC) light curves. An alert was then issued by the \textit{TESS} Objects of Interest (TOI) working group (TOI~495.01) based on the detected periodic transit signal.

\begin{figure*}[!htbp]
    \centering
    \includegraphics[width=1\textwidth]{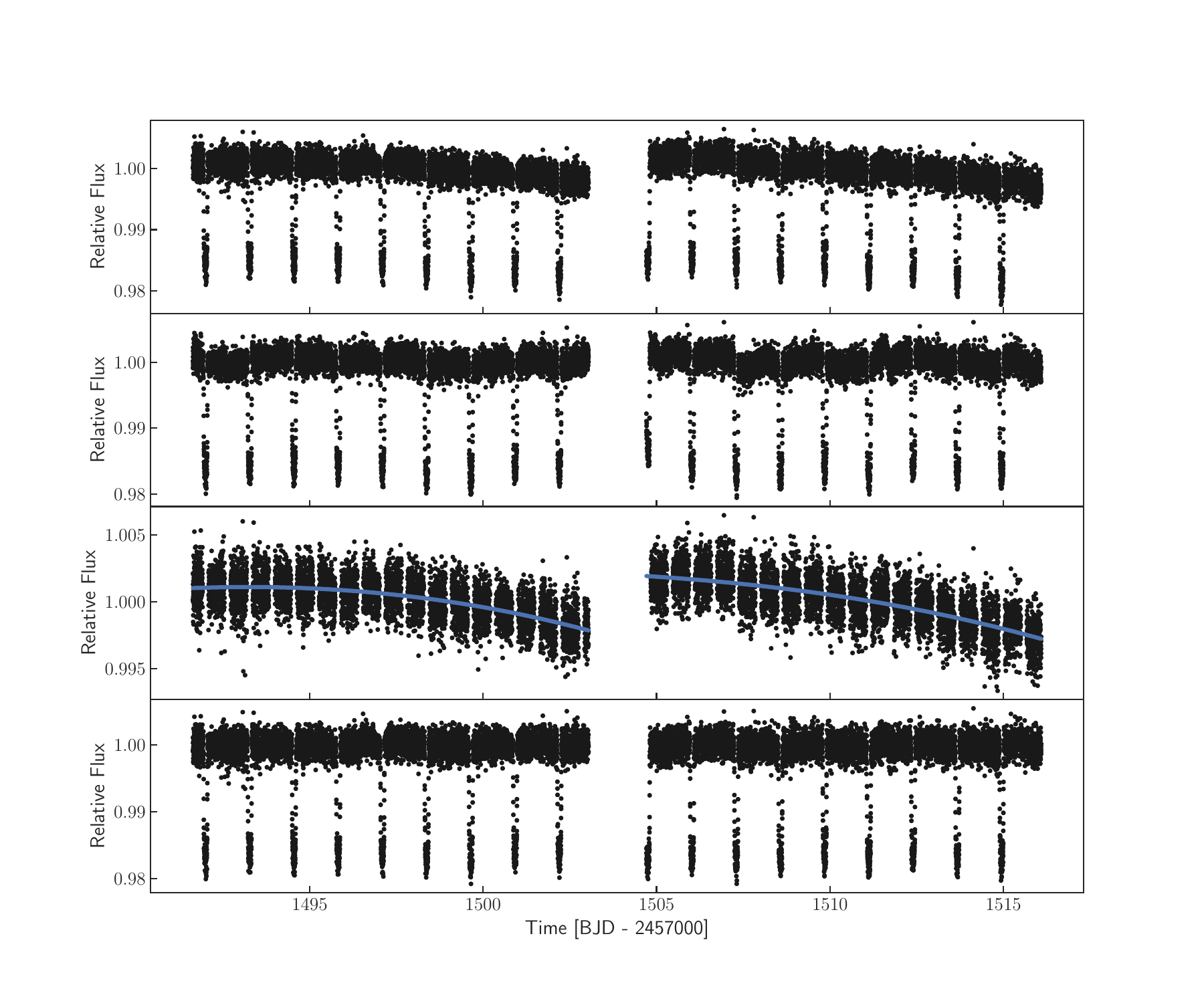}
    \caption{Top: the SAP \textit{TESS} light curve of WASP-121. Second from the top: the PDC light curve of WASP-121, showing indications of excessive red noise. Third from the top: SAP light curve of WASP-121 after masking both the primary transits and the secondary eclipses of WASP-121\,b (black) and the spline model (blue). Bottom: the detrended light curve on which the later analyses are based. In all panels, there is a gap in the middle of the data set due to \textit{TESS} pausing data collection around perigee.}
    \label{figr:lcur_totl}
\end{figure*}

In order to produce the SAP light curve, the SPOC pipeline subtracts a background photon count estimate from the total number of photons inside a given aperture, whereas the PDC data product also contains a correction for red noise as well as the decrease in the transit depth due to known nearby sources. These initial data are shown in the first (SAP) and second (PDC) panels of Figure~\ref{figr:lcur_totl}, respectively. We inspected both data sets and concluded that the PDC light curve was not suitable for further analysis due to residual systematics in the data at short time scales, especially prominent in the second orbit of Sector 7. Red noise is mostly in the form of discontinuities and systematic trends introduced by instrumental effects such as changes in the thermal state of \textit{TESS} and pointing instabilities. Therefore, we decided to fit a cubic spline to the SAP data collected during each individual orbit after masking out the flagged data points as well as the transits as shown in the third panel of Figure~\ref{figr:lcur_totl}. This procedure filtered systematics introduced at a time scale much longer than the orbital period, while preserving the phase curve modulations.

When fitting cubic splines, we increased the number of knots until the residual per degree of freedom fell below unity. The resulting coefficients and knot locations are given in Table~\ref{tabl:coef} for each orbit separately. We then based the subsequent analyses on the detrended light curve obtained by removing the spline fit from the SAP light curve. The resulting detrended light curve is shown in the bottom panel of Figure~\ref{figr:lcur_totl}. It is true that this leaves some red noise at short time scales in the resulting light curve, which can potentially bias our nominal fit. As a crosscheck, this detrended light curve was also modeled using a Gaussian Process (GP) in an alternative analysis, which will be described in Section~\ref{sect:meth}.
\begin{table}[]
    \centering
    \begin{tabular}{c|c|c}
        &  First orbit & Second orbit \\
        \hline
        \hline
        $T_{i}$ [BJD] & 2458491.63452417 & 2458503.0401359 \\
        \hline
        $T_{f}$ [BJD] & 2458504.83318593 & 2458516.08867022 \\
        \hline
        $\beta_0$ & 0.00103486 & 1.91152409e-03 \\
        \hline
        $\beta_1$ & 0.00140942 & 1.28100318e-03 \\
        \hline
        $\beta_2$ & 0.00048103 & -5.56292798e-05 \\
        \hline
        $\beta_3$ & -0.00212276 & -2.75337528e-03 
    \end{tabular}
    \caption{Table of cubic spline coefficients, $\beta_i$ for $i~=~\{0,1,2,3\}$, for the first and second orbits, which begin and end at $T_i$ and $T_f$.}
    \label{tabl:coef}
\end{table}

When extracting the light curve of a target, transits are diluted (i.e., their depths are decreased) due to light contamination from neighboring stars into the photometric aperture. Given the 21\arcsec pixel scale of \textit{TESS} and the bright neighbors within $\sim$100\arcsec of WASP-121, WASP-121's \textit{TESS} light curve is expected to be diluted by $\sim$8.3\%. However, the amount of dilution cannot be known precisely due to uncertainties in the point spread function. This results in an uncertainty that needs to be propagated forward during inference. Hence, in order to take into account the dilution of the transits in the (uncorrected) SAP light curves, we included a dilution parameter in the light curve model, imposing a Gaussian prior centered on the expected value of 8.3\%.

The available data on WASP-121\,b include: the secondary eclipse data at 3.6 $\mu$m and 4.5 $\mu$m collected by \textit{Spitzer} \citep{Garhart+2019}, $z^\prime$ photometry measured by \textit{TRAPPIST} \citep{Delrez+2016}, $K_s$ band photometry \citep{KovacsKovacs2019}, G102 and G141 \textit{HST}/WFC3 emission spectroscopy \citep{Evans+2017, Mikal-Evans+2019}; the transmission spectroscopy data collected during the primary transit from \citep{Evans+2018}; and the \textit{TESS} phase curve analyzed in this work. We utilize the secondary eclipse data to perform an atmospheric retrieval (see Section~\ref{sect:modlatmo}).

\section{Methods}
\label{sect:meth}

In this section, we lay out our methodology for modeling
\begin{enumerate}
    \item the \textit{TESS} light curve to infer the orbital properties and the phase curve characteristics of the WASP-121 system,
    \item the resulting phase curve characteristics to infer the temperature distribution on WASP-121\,b,
    \item the multiband spectrum of WASP-121\,b's dayside to retrieve the atmospheric properties of the dayside of WASP-121\,b.
\end{enumerate}

\subsection{Phase curve}
\label{sect:modl}

The variation of the brightness of a star-exoplanet system during orbit, i.e., a phase curve, can be broken down into several components. First, the gravitational field of the star and the exoplanet tidally distorts the two bodies according to the Roche potential, causing ellipsoidal variations in the brightness of the system at a period equal to the half of the orbital period. Second, the radial velocity of the orbiting planet results in the redward and blueward shifting (i.e., Doppler boosting) of light from the star, at the orbital period. Lastly, the thermal and reflected emission from the planet change during an orbit, as a result of the changing fraction of the illuminated and heated surface of the planet. We will refer to the sum of the thermal and reflected emission from the planet as the \emph{planetary modulation} component.

\subsubsection{Ellipsoidal variation}

Ellipsoidal variation is a measure of how the tidal gravitational field of the companion compares to the surface gravity of the host star. Its fractional amplitude can be expressed to leading order as \citep{Morris1985, MorrisNaftilan1993}
\begin{equation}
    D_{\rm e} = \alpha_{\rm e} \dfrac{M_{\rm p}}{M_\star} \sin^2 i \Bigg(\dfrac{R_\star}{a}\Bigg)^3,
\end{equation}
where $\alpha_{\rm e}$ is a constant that depends on the linear limb darkening and gravity darkening coefficients, $M_{\rm p}$ and $M_\star$ are the mass of the planet and the star, $i$ is the orbital inclination, $R_\star$ is the radius of the star and $a$ is the semi-major axis. 

We used the limb darkening and gravity darkening coefficients in \citet{Claret2017} to predict the ellipsoidal variation amplitude of the WASP-121 system as $\sim$20 ppm, which justified the inclusion of ellipsoidal variations into our light curve model. Although strong ellipsoidal variations can manifest higher order terms (i.e., harmonics), we only included the first order (i.e., fundamental) term into the model and neglected higher order terms.

\subsubsection{Doppler boosting}

During the orbit of the companion, the line-of-sight velocity of the host star varies periodically, causing photons to be blue or red shifted. Therefore, although the bolometric flux does not change, the brightness of the host in a particular passband varies at the orbital period with the fractional amplitude of \citep{Shporer2017}

\begin{equation}
    D_{\rm b} = 4 \alpha_{\rm b} \frac{K}{c},
\end{equation}
where
\begin{equation}
    \alpha_{\rm b} = \int \frac{1}{4} \frac{xe^x}{e^x-1}dx, x \equiv \frac{hc}{kT\lambda}.
\end{equation}
Here, $K$ is the radial velocity semi-amplitude, $c$ is the speed of light, $h$ is the Planck constant, $k$ is the Boltzmann constant, $T$ is the temperature of the star, and $\lambda$ is the wavelength. The integral is taken inside the TESS passband as shown in the top row of Figure~\ref{figr:spec}. The Doppler boosting fractional amplitude we expect for this system was estimated as $\sim$1 ppm, which is much smaller than the uncertainties of the photometric data. Therefore, we did not include Doppler boosting in our light curve model.

\subsubsection{Thermal and reflected planetary emission}

As the planet orbits the host star, a changing fraction of its surface is illuminated. Therefore, both the reflected and thermal emission from the orbiting planet is expected to follow roughly a cosine at the orbital period whose maximum occurs near phase 0.5.

At the superior conjunction, the total secondary depth,
\begin{equation}
    D_{\rm s} = D_{\rm t} + D_{\rm r},
    \label{equa:deptseco}
\end{equation}
can be generically expressed as the sum of a thermal, $D_{\rm t}$, and reflected, $D_{\rm r}$, components. Equation~\ref{equa:deptseco} shows that the thermal and reflected emissions are degenerate given some secondary eclipse data in a single band. Having spectral data, therefore, becomes crucial for placing an independent constraint on the thermal emission in order to break this degeneracy \citep{Kreidberg+2018, Mansfield+2018}.

The thermal component can further be written as \citep{Charbonneau+2005}
\begin{equation}
    D_{\rm t} = \Bigg(\frac{R_{\rm p}}{R_\star}\Bigg)^2 \times \dfrac{\int \mathcal{T}(\lambda) F_{\rm p}(\lambda,T_{\rm p}) d\lambda}{\int \mathcal{T}(\lambda) F_\star(\lambda,T_\star) d\lambda},
\end{equation}
where $R_{\rm p}$ and $R_\star$ are the planetary and stellar radii, $\mathcal{T}$ is the \textit{TESS} throughput and $F_{\rm p}(\lambda,T)$ and $F_\star(\lambda,T)$ are the planetary and stellar fluxes, which are roughly blackbodies with effective temperatures $T_{\rm p}$ and $T_\star$. The reflected component, on the other hand, is given by \citep{Rodler+2010}
\begin{equation}
    D_{\rm r} = A_{\rm g} \Bigg(\frac{R_{\rm p}}{a}\Bigg)^2,
\end{equation}
where $A_{\rm g}$ is the geometric albedo in the \textit{TESS} passband, which quantifies the reflectivity of a planet relative to that of a Lambertian disk and differs from the Bond albedo, $A_{\rm B}$, which is the ratio of the reflected bolometric flux to that received from the host star. 

The brightness of an ultra-hot Jupiter in the infrared is dominated by its thermal emission. However, it may get an increasing contribution from reflection at shorter wavelengths, such as those probed by \textit{TESS}. Reflection of incident light causes the equivalent temperature of the planet to be reduced at fixed stellar radius and orbital semi-major axis. Another process with which the dayside equivalent temperature of a planet can be reduced is heat recirculation from the dayside to the nightside, which can partially homogenize the temperature distribution across the exoplanet.

\subsubsection{Light curve model}

We modeled the \textit{TESS} light curve using \texttt{allesfitter} \citep{GuentherDaylan2019a, GuentherDaylan2019b}. Given the orbital phase between 0 and 1, $\phi$, we assume the primary transit and secondary eclipse to happen at phases 0 and 0.5, respectively. The resulting light curve model can be expressed as a linear combination of the following components:
\begin{itemize}
\item Baseline stellar emission. This component assumes a constant stellar emission, which is partially occulted during the primary transit of the planet according to our transit model (calculated via \texttt{ellc}),
\item Constant planetary emission. This component models the planet as a constant source of emission. It is parametrized by the surface brightness ratio of the planet and the star, $J$, outside the secondary eclipse and is 0 otherwise (calculated via \texttt{ellc}).
\item Planetary modulation. The modulation has the form of $D_{\rm p} (1 - \cos 2\pi(\phi + \Delta_{\rm p}/P))/2$ outside the secondary eclipse and is 0 otherwise. Here, $P$ is the orbital period and $\Delta_{\rm p}$ allows us to model a nonzero phase shift.
\item Ellipsoidal variation. The ellipsoidal variation is modeled as $D_{\rm e} (1 - \cos 4\pi\phi)/2$, i.e., a sinusoidal at half of the orbital period that peaks at quadrature (0.25 and 0.75) phases.
\item A constant offset to absorb any normalization bias denoted by $O$.
\end{itemize}

We model the baseline stellar emission, primary transit, and the secondary eclipse in terms of the following parameters: the planet-to-star radius ratio, $R_{\rm p}/R_\star$; the ratio of the sum of the stellar and planetary radius and the semi-major axis of the orbit, $(R_\star + R_{\rm p})/a$; the cosine of the orbital inclination, $\cos i$; the epoch, $T_{0}$; the period, $P$; the terms parametrizing the orbital eccentricity, $\sqrt{e}\cos \omega$ and $\sqrt{e}\sin \omega$, where $e$ and $\omega$ are the eccentricity and argument of periapsis; the dilution in the \textit{TESS} frames, $D_{\rm TESS}$; transformed limb darkening coefficients $q_{1}$ and $q_{2}$; the logarithm of the average measurement uncertainty of the relative flux data, $\log \sigma$; and a flat baseline fitted to the light curve, $O$. We assume zero eccentricity \citet{Delrez+2016}. As for limb darkening, we use the transformed set of limb darkening coefficients, $q_1 \equiv (u_1 + u_2)^2$ and $q_2 \equiv 0.5 (u_1 + u_2)^{-1}$ \citep{Kipping2013}, which allows us to efficiently sample from the posterior of the limb darkening coefficients. When fitting the relative flux data, we allow a free rescaling of the uncertainties, where we first normalize the measurement uncertainties of relative flux by dividing them by the mean uncertainty and then multiplying these \emph{normalized} uncertainties by $\sigma$ defined above.

The main challenge in modeling data with systematic errors is red noise, which is the result of all instrumental and astrophysical processes that are not available in the forward-model used to fit a given data set. Thanks to their flexibility, GPs can be used to model red noise. In particular, \texttt{allesfitter} has a built-in GP with the Mat\'ern 3/2 kernel with the hyperparameters $\sigma_{\rm GP}$ (amplitude) and $\rho_{\rm GP}$ (time scale). In this work, our nominal fit employs a light curve model with a flat baseline. However, as a crosscheck to our nominal fit, we performed an additional fit using a GP as the baseline model.

We note that the \textit{TESS} data, being bluer than other available secondary eclipse data on the system, probe relatively higher altitudes in the atmosphere, and thus has more constraining power on the temperature inversion. It is also the case that, being bluer than other passbands, the \textit{TESS} data are more susceptible to stellar variability. Our light curve model with the GP can absorb any stellar variability. Therefore, this alternative fit also brackets the range of solutions consistent with the \textit{TESS} data, marginalizing over stellar variability.

Given the above-mentioned forward model and the spline-detrended \textit{TESS} light curve, we sample from the posterior distribution subject to certain priors using \texttt{emcee} \citep{Foreman-Mackey2013}. In order to make sure that the samples taken from the posterior are fair and that the chains are converged, we collect samples until the chain length is at least 30 times that of the integrated autocorrelation time of all parameters. The definition of our parameters, the priors imposed, and the posterior are listed in Table~\ref{tabl:resualle}.

\subsection{Temperature distribution}

In the absence of any heat circulation, the equilibrium temperature of the planet, $T_{\rm eq}$, is given by
\begin{equation}
    T_{\rm eq} = T_\star (1 - A_{\rm B})^{1/4} \sqrt{\dfrac{R_\star}{2a}}
    \label{equa:tmptsmaj}
\end{equation}
where $T_\star$ is the effective temperature of the host star, $a$ is the orbital semi-major axis, $E$ is the planetary emissivity and $A_{\rm B}$ is the Bond albedo. Furthermore, when there is heat recirculation between the day and night sides of a planet, the dayside and the nightside temperatures, $T_{\rm d}$ and $T_{\rm n}$, can be parametrized as, 
\begin{align}
    T_{\rm d} &= T_{\rm eq} \Big(\frac{2}{3} - \frac{5}{12} \varepsilon \Big)^{1/4} \equiv \psi T_{d,0}, \\
    T_{\rm n} &= T_{\rm eq} \Big(\frac{\varepsilon}{4}\Big)^{1/4},
    \label{equa:tmptdayy}
\end{align}
respectively, where $\varepsilon$ parametrizes the heat recirculation efficiency \citep{CowanAgol2011}. An $\varepsilon$ of 0 indicates no heat recirculation whereas an $\varepsilon$ of 1 implies instantaneous redistribution of incident energy over the planet. Here, the dayside temperature in the case of perfect emissivity and the absence of any heat recirculation, reflection and internal heat is denoted by $T_{d,0}$. Therefore $\psi$ is a parameter that takes into account the effect of any reflection, heat recirculation and planetary internal heat. In particular, $\psi$ is equal to 1 when there is no heat recirculation, the planet does not reflect any light, and in the limit of zero internal heat.

In the absence of any heat recirculation, the point on the surface of a tidally-locked planet directly pointing to the host star, i.e., the substellar point, is expected to be the hottest point on the surface of the planet \citep{Mazeh2008}. However, heat recirculation in the atmosphere (e.g., via winds due to atmospheric pressure gradients) can transport energy and result in an eastward phase shift. Another potential source of phase shift is reflective clouds on the western hemisphere. Although the dayside temperature of ultra-hot Jupiters is typically too high for reflective condensates to be sustained, refractory species can still form on the nightside of the planet. Furthermore, the western terminator of a hot Jupiter is expected to be colder than the eastern one, because the super-rotating equatorial jet moves cold gas from the nightside to the western terminator of the planet. When this gas is sufficiently cool, condensates can remain stable on the western terminator. This can result in some reflected light from the western limb and cause a westward phase shift \citep{Parmentier+2016}. Therefore, there can be distinct phase shifts for the thermal and reflected emissions.

In our nominal fit, we use a light curve model with a single planetary modulation component with amplitude $D_{\rm p}$ and phase shift $\Delta_{\rm p}$. This effectively accounts for the sum of the reflected and thermal components, which are largely degenerate. We do, however, perform an alternative fit, where we use two different planetary modulation components with amplitudes $D_{\rm 1}$ and $D_{\rm 2}$ and phase shifts $\Delta_{\rm 1}$ and $\Delta_{\rm 2}$. Nevertheless, we present this fit only as a crosscheck, as it potentially overfits the data, since the two components are largely degenerate.

Given an estimate of the thermal phase shift and the dayside and nightside temperatures of a planet, one can also parametrically infer the temperature distribution. Towards this purpose, we solve the kinematic differential equation \citep{ZhangShowman2017}
\begin{equation}
\frac{\partial T}{\partial t} + \frac{1}{\tau_{adv}} \frac{\partial T}{\partial \lambda} = \frac{1}{\tau_{rad}} (T_{eq}(\lambda) - T)
\label{equa:tmpt}
\end{equation}
which describes the heat recirculation, and hence, the temperature distribution on the planet, where $T$ is the effective temperature, $\lambda$ is the longitude, $\tau_{adv}$ and $\tau_{rad}$ are the advective and radiative time scales, respectively. Their ratio, $\xi \equiv \tau_{rad}/\tau_{adv}$, is related to, but distinct from the heat recirculation efficiency $\varepsilon$ defined previously. Although they are both zero in the limit of vanishing heat recirculation, the limit $\xi \to \infty$ corresponds to $\varepsilon$ $\sim$0.95. At constant longitude, we assume that the temperature on the planet is proportional to $\cos(\theta)$ as a function latitude, $\theta$. 
The solution of Equation~\ref{equa:tmpt} yields the form
\begin{equation}
T(\lambda) = 
\begin{cases}
T_{\rm n} + T_1 \cos \lambda_s \cos(\lambda - \lambda_s) + \eta T_1 e^{-\lambda/\xi} \\ \hspace{6em}\text{ for $-\pi/2 < \lambda < \pi/2$}
\\
T_{\rm n} + \eta T_1 e^{-(\pi+\lambda)/\xi} \text{ for } -\pi < \lambda < -\pi/2 \\
T_{\rm n} + \eta T_1 e^{(\pi-\lambda)/\xi}  \text{ for } \pi/2 < \lambda < \pi,
\end{cases} 
\end{equation}
where $T_1$ is the temperature difference between the anti-stellar and sub-stellar points on the planet, $T_{\rm n}$ is the temperature of the nightside, and $\eta$ and $\lambda_s$ are defined as

\begin{equation}
\eta = \frac{\xi}{1+\xi^2} \frac{
1}{e^{\pi/2\xi} -1} \Bigg(e^{\frac{
\pi}{2\xi}} + e^{\frac{
3\pi}{2\xi}} \Bigg)
\end{equation}
and
\begin{equation}
\lambda_s = \tan^{-1}(\xi).
\end{equation}

Given a zonal jet that recirculates heat, and hence, causes a nonzero thermal phase shift, $\lambda_m$, $\xi$ can be calculated using
\begin{equation}
\sin(\lambda_s - \lambda_m) e^{\lambda_m/\xi} = \frac{\eta}{\xi \cos \lambda_s}
\end{equation}
using the median thermal phase shift inferred from the \textit{TESS} data. This allows an estimate of the temperature distribution on the planet as shown in Figure~\ref{figr:tmpt}, if the phase shift had originated solely from heat recirculation.

\subsection{Atmospheric modeling}
\label{sect:modlatmo}

We model the atmosphere of WASP-121\,b using a one dimensional radiative transfer model. We use ATMO \citep{Amundsen2014, Tremblin+2015, Tremblin+2016, Tremblin+2017, Drummond+2016, Goyal+2017}, which solves the radiative transfer equation on a one-dimensional grid assuming hydrostatic and radiative-convective equilibrium. ATMO assumes \emph{double-gray equilibrium}, where the opacity of the atmosphere is assumed to be constant in the optical and infrared bands. These two opacities are modeled as free and independent. This simplification enables analytic solutions of the radiative transfer equation, which otherwise requires a slow, numerical solution. This reduces the computational expense of forward-modeling to $\sim$1 sec as opposed to tens of minutes, enabling samples from the posterior of the generative model to be taken in a computationally efficient way.

ATMO infers the properties of an atmosphere subject to irradiation consistent with a measured emission spectrum, by sampling from a radiative transfer model with 6 free parameters. These are the irradiation efficiency, $\psi$, base-10 logarithm of the relative metallicity (i.e., abundance of all elements except hydrogen, carbon and oxygen), [M/H], infrared opacity, $\kappa_{IR}$, the optical to infrared opacity ratio, $\gamma$, base-10 logarithm of the relative carbon abundance, [C/H], and base-10 logarithm of the relative oxygen abundance, [O/H].

Given the input optical and infrared opacities, ATMO forward-models the pressure-temperature profile assuming radiative-convective equilibrium. The resulting pressure-temperature profile is used together with the input elemental abundances to solve for the atmospheric composition assuming chemical equilibrium. Finally, the emission spectrum is calculated using the spectral line lists of the molecules in the atmosphere. The latter is done via the \emph{correlated-k} method, i.e., using precomputed tables of line lists to make it computationally efficient. For this retrieval, we included the opacity sources of H$_2$, He, H$_2$O, CO$_2$, CH$_4$, NH$_3$, Na, K, Li, Rb, Cs, TiO, VO, FeH, PH$_3$, H$_2$S, HCN, C$_2$H$_2$, SO$_2$, Fe, and H$^-$. We note that this is a \emph{one-way self-consistent} method. That is, the chemistry is computed self-consistently given the pressure-temperature profile. But the pressure-temperature profile is not necessarily consistent with the resulting chemistry.

Thermal dissociation is an important process in the atmospheres of ultra-hot Jupiters, which significantly reduces the abundance of, and introduces a vertical abundance gradient for relatively loosely bound chemical species such as H$_2$O at high altitudes \citep{Arcangeli+2018, Arcangeli+2019, Parmentier+2018}. In particular, WASP-121\,b contains a significant H$_2$O emission feature, indicating the H2O at it's photosphere is not fully dissociated \citep{Evans+2017}. Therefore, ATMO also takes into account thermal dissociation of chemical species. Given the potential for significant reflected light in the \textit{TESS} passband, we also included isotropic scattering.

In addition to the irradiation from the host star, another source of energy for the atmosphere is that of internal heat. In this work, we fix the internal temperature to a fiducial value of 100 K, as it would be highly degenerate with other sources with the available data.

\section{Results}
\label{sect:resu}

In this section, we first discuss our results regarding the analysis of the \textit{TESS} light curve of WASP-121. Then, we discuss the results of running ATMO based on the inferred characteristics of the phase curve of WASP-121\,b.

\subsection{Phase curve characteristics}

We performed three light curve fits with different configurations. All of our results are shown in Table~\ref{tabl:resualle}. The \emph{nominal} fit incorporates a flat baseline and a single planetary modulation component (second column). An alternative fit with a GP baseline and single planetary modulation component is shown in the third column. Finally, another alternative fit with a flat baseline, but two planetary modulation components is shown in the fourth column. The table also shows our priors for each parameter.

We refer to the fit with flat baseline and single planetary modulation component as our \emph{nominal} fit. This is because both the two-component fit and the GP baseline can potentially overfit the data and are only intended as crosschecks. The results with a flat baseline and a GP baseline are shown to be statistically consistent. In the two component fit, neither of the phase shifts are found to be statistically significant. In the remainder of this paper, we use the results of the nominal fit for further analyses, i.e., the inference of the temperature distribution and the atmospheric retrieval.

\begin{table*}[]
    \centering
    \caption{Parameters, posterior quantiles, and priors of the fitted and derived parameters of the global light curve model. The short hand notations FBSC, FBDC, and GBSC refer to Flat Baseline Single planetary Component, Flat Baseline Double planetary Component, and Gaussian process baseline Single planetary Component, respectively.\vspace{-1em}}
    \begin{tabular}{m{0.27\textwidth}|ccc|c}
Parameter                    &         & Posterior                  &  & Prior  \\
\hline
\hline
\hline
                             &  FBSC (Nominal fit)  & FBDC  (Alternate fit)                & GBSC (Alternate fit) & \\
\hline
\hline
Fitting parameters & & & & \\
\hline
\hline
$R_{p} / R_\star$                                   & $0.12488\pm0.00072$          & $0.12488\pm0.00072$                  & $0.12486\pm0.00072$                  & U [0,1] \\
$(R_\star + R_{p}) / a$                         & $0.3061\pm0.0029$            & $0.3062\pm0.0029$                    & $0.3059\pm0.0029$                    & U [0,1] \\
Cosine of orbital inclination; $\cos{i}$        & $0.0825_{-0.011}^{+0.0096}$  & $0.0826_{-0.011}^{+0.0097}$          & $0.0818_{-0.011}^{+0.0096}$          & U [0,1] \\
Epoch; $T_{0}$ [BJD-2458000]                      & $504.748003\pm0.000064$      & $504.748004\pm0.000064$              & $504.747996\pm0.000063$              & U [504,505] \\
Orbital period; $P$ [day]                       & $1.274928\pm0.000011$        & $1.274928\pm0.000011$                & $1.274929\pm0.000011$                & U [1,2] \\
$\sqrt{e} \cos{\omega}$                         & 0                            & 0                                    & 0                                    & fixed \\
$\sqrt{e} \sin{\omega}$                         & 0                            & 0                                    & 0                                    & fixed \\
Limb darkening; $q_{1}$                             & $0.115_{-0.035}^{+0.044}$    & $0.115_{-0.035}^{+0.044}$            & $0.115_{-0.035}^{+0.043}$            & U [0,1] \\
Limb darkening; $q_{2}$                             & $0.42_{-0.13}^{+0.18}$       & $0.42_{-0.13}^{+0.18}$               & $0.42_{-0.13}^{+0.18}$               & U [0,1] \\
Error scale factor; $\log{\sigma}$                  & $-6.7616\pm0.0055$           & $-6.7616\pm0.0056$                   & $-6.7763\pm0.0055$                   & U [-9,-3] \\
Surface brightness ratio; $J$                       & $0.0040_{-0.0022}^{+0.0025}$ & $0.0032_{-0.0021}^{+0.0025}$         &  $0.0050_{-0.0032}^{+0.0041}$        & U [0,1] \\
Beaming amplitude; $D_\mathrm{b}$                   & 0.                           & 0                                    & 0                                    & fixed \\
EV amplitude; $D_\mathrm{e}$                        & $0.0083_{-0.0061}^{+0.012}$  & $0.0080_{-0.0060}^{+0.012}$          & $0.0094_{-0.0069}^{+0.014}$          & U [0,10] \\
PM amplitude; $D_\mathrm{p}$                        & $0.418\pm0.034$              &                                      & $0.412_{-0.061}^{+0.054}$            & U [0,10] \\
PM phase shift; $\Delta_\mathrm{p}$                 & $-0.022\pm0.014$             &                                      & $-0.018\pm0.032$                     & U [-0.5,0.5] \\
PM amplitude (c1); $D_\mathrm{1}$          &                              & $0.127_{-0.091}^{+0.16}$             &                                      & U [0,10] \\
PM phase shift (c1); $\Delta_\mathrm{1}$ [day]   &                              & $-0.03\pm0.17$                       &                                      & U [-0.5,0.5] \\
PM amplitude (c2); $D_\mathrm{2}$           &                              & $0.326_{-0.18}^{+0.094}$             &                                      & U [0,10] \\
PM phase shift (c2); $\Delta_\mathrm{2}$ [day]   &                              & $-0.018_{-0.043}^{+0.055}$           &                                      & U [-0.5,0.5] \\
Dilution; $D_{\rm TESS}$                                       & $0.0765\pm0.0082$            & $0.0766\pm0.0082$                    & $0.0763\pm0.0083$                    & G [0.083,0.0083] \\
Additive offset; $O$                                & $-0.000190\pm0.000019$       & $-0.000202_{-0.000026}^{+0.000022}$  &                                      & U [-1,1] \\
$\log{\sigma_{\rm GP}}$                                 &                              &                                      & $-6.04_{-0.18}^{+0.19}$              & U [-10,0] \\
$\log{\rho_{\rm GP}}$                                   &                              &                                      & $1.052_{-0.040}^{+0.089}$            & U [1,2] \\
\hline
\hline
Derived parameters & & & & \\
\hline
\hline
$R_\star/a_\mathrm{p}$                                & $0.2722\pm0.0025$              & $0.2722\pm0.0025$              & $0.2720\pm0.0025$  \\
$a_\mathrm{p}/R_\star$                                & $3.674\pm0.035$                & $3.674\pm0.035$                & $3.677\pm0.035$ \\
$R_\mathrm{p}/a_\mathrm{p}$                           & $0.03399\pm0.00042$            & $0.03399\pm0.00042$            & $0.03396\pm0.00042$ \\
Planet radius; $R_\mathrm{p}$ [$\mathrm{R_{J}}$]      & $1.772\pm0.038$                & $1.772\pm0.038$                & $1.771\pm0.038$ \\
Semi-major axis; $a_\mathrm{p}$ [AU]                  & $0.02491\pm0.00057$            & $0.02491\pm0.00058$            & $0.02493\pm0.00057$ \\
Orbital inclination; $i_\mathrm{p}$ [degree]          & $85.26_{-0.56}^{+0.64}$        & $85.26_{-0.56}^{+0.64}$        & $85.31_{-0.55}^{+0.65}$ \\
Impact parameter; $b_\mathrm{\rm tr}$                  & $0.303_{-0.038}^{+0.033}$      & $0.303_{-0.039}^{+0.033}$      & $0.301_{-0.039}^{+0.032}$ \\
Total transit duration; $T_\mathrm{t}$ [h]        & $2.9250\pm0.0082$              & $2.9249\pm0.0082$              & $2.9246\pm0.0081$ \\
Full transit duration; $T_\mathrm{f}$ [h]        & $2.202\pm0.016$                & $2.202\pm0.016$                & $2.203\pm0.016$  \\
Planet eq. temperature; $T_\mathrm{eq}$ [K]           & $2179\pm49$                    & $2179\pm49$                    & $2179\pm48$ \\
Transit depth; $\delta_{\rm tr}$ [ppt]                & $17.12\pm0.17$                 & $17.13\pm0.17$                 & $17.12\pm0.17$ \\
\hline
Limb darkening; $u_\mathrm{1}$                        & $0.285\pm0.058$                & $0.285\pm0.058$                & $0.285\pm0.057$ \\
Limb darkening; $u_\mathrm{2}$                        & $0.06\pm0.11$                  & $0.05\pm0.11$                  & $0.05\pm0.11$\\
Host density; $\rho_\mathrm{\star}$ [cgs]             & $0.577\pm0.016$                & $0.577\pm0.016$                & $0.579_{-0.016}^{+0.017}$  \\
\hline
Secondary eclipse depth; $\delta_\mathrm{s}$ [ppt]    & $0.482_{-0.041}^{+0.039}$      & $0.484\pm0.039$                & $0.491_{-0.039}^{+0.043}$ \\
Nightside flux [ppt]; $D_\mathrm{n}$ [ppt]            & $0.065_{-0.037}^{+0.040}$      & $0.050_{-0.031}^{+0.041}$      & $0.082_{-0.052}^{+0.061}$ \\
Dayside temperature; $T_{\rm d}$ [K]                  & $3012\substack{+40 \\ -42}$    & $3029\substack{+46 \\ -48}$    & $3023_{+44}^{-41}$ \\
Nightside temperature; $T_{\rm n}$ [K]                & $2022\substack{+254 \\ -602}$  & $1907\substack{+309 \\ -579}$  & $2016_{+639}^{-344}$ \\
Temperature contrast; $\delta_{\rm T}$                & $0.33\substack{+0.20 \\ -0.08}$& $0.37\substack{+0.19 \\ -0.10}$& $0.33\substack{+0.21 \\ -0.12}$ \\
\hline 
    \end{tabular}
    \begin{minipage}{16cm} 
    The equilibrium temperature is reported assuming a Bond albedo of 0.3. PM stands for planetary modulation. c1 and c2 denote the two planetary modulation components with independent amplitudes and phase shifts.
    \end{minipage}

    \label{tabl:resualle}
\end{table*}

\begin{figure*}
    \centering
    \includegraphics[width=0.95\textwidth]{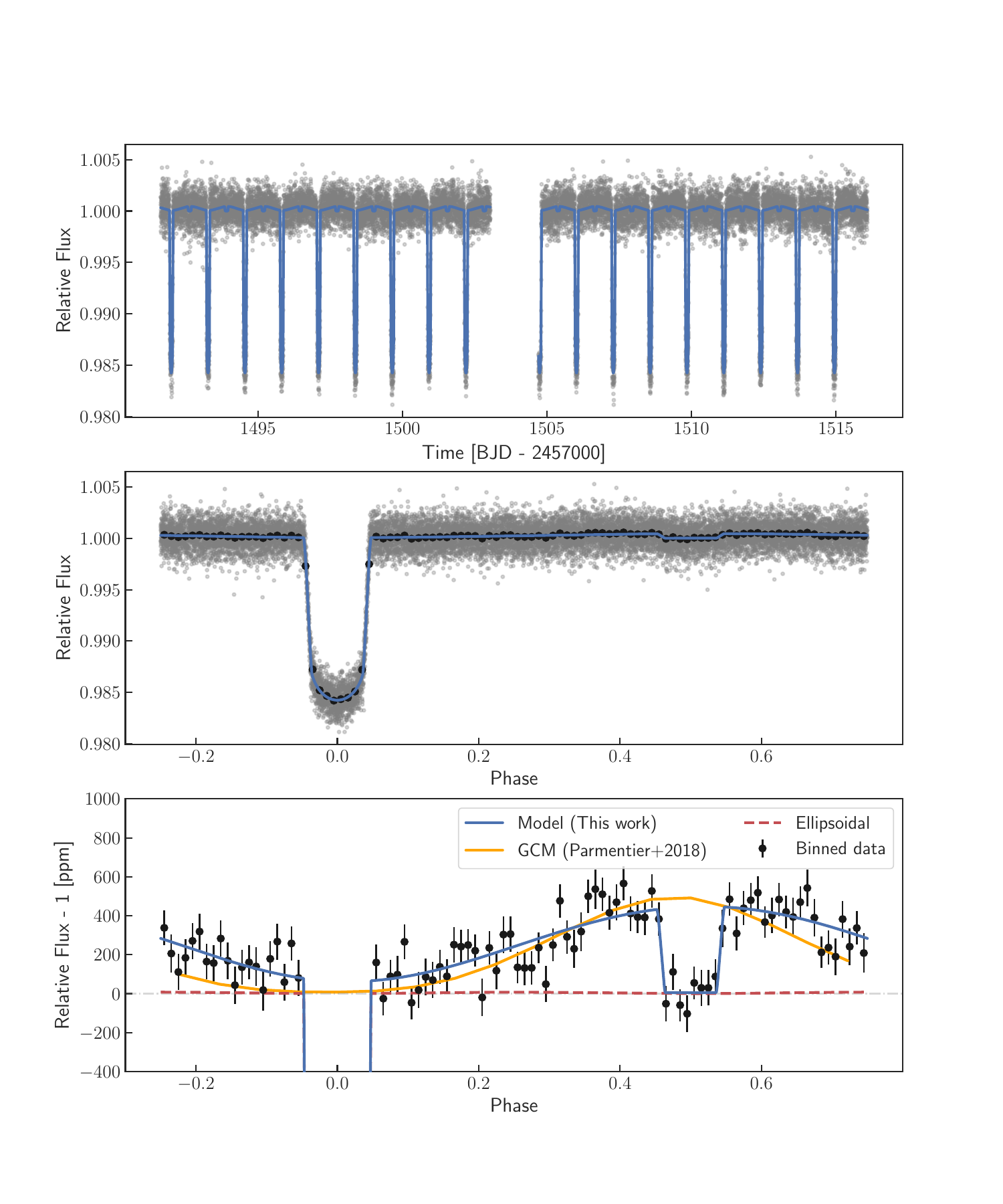}
    \caption{Top: the detrended \textit{TESS} light curve of WASP-121 (grey) and the median model light curve drawn from the posterior (blue). Middle: The \textit{TESS} phase curve of WASP-121\,b. The grey points denote the raw light curve. Black points show the binned light curve. Blue line is the posterior median phase curve. Bottom: The \textit{TESS} phase curve zoomed around unity showing the binned phase curve (black), the posterior median model (blue), ellipsoidal modulation (red dashed line), and, for comparison, a Global Circulation Model (GCM) prediction \citep{Parmentier+2018} (yellow line) (see Section \ref{sect:disc}). The grey centerline highlights zero, i.e., no emission from WASP-121\,b.}
    \label{figr:pcur_alle}
\end{figure*}

\begin{figure*}[!htbp]
    \centering
    \includegraphics[width=1\textwidth]{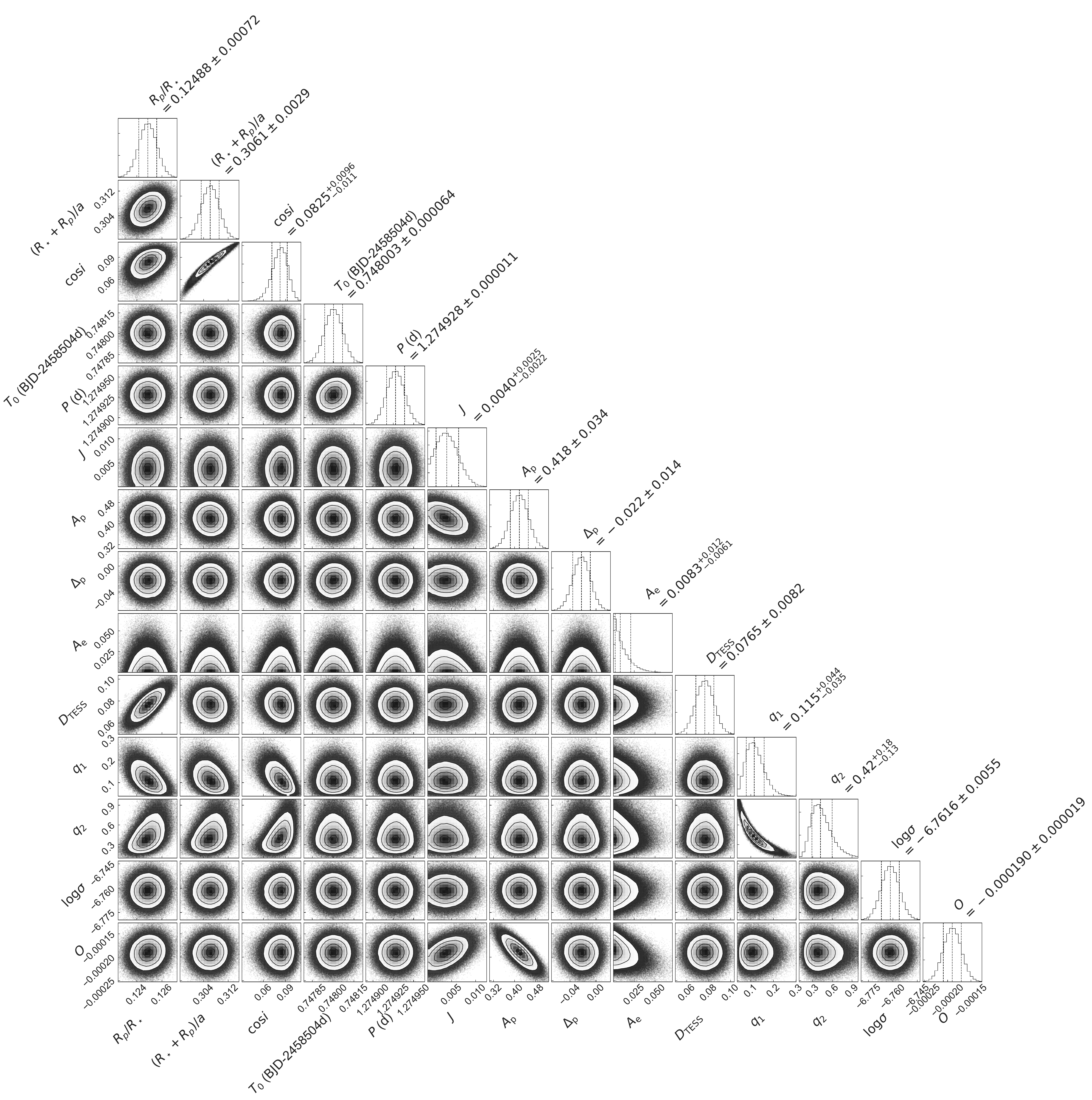}
    \caption{Posterior probability distribution of the light curve model parameters.}
    \label{figr:postalle}
\end{figure*}

\begin{figure*}[!htbp]
    \centering
    \includegraphics[width=0.95\textwidth]{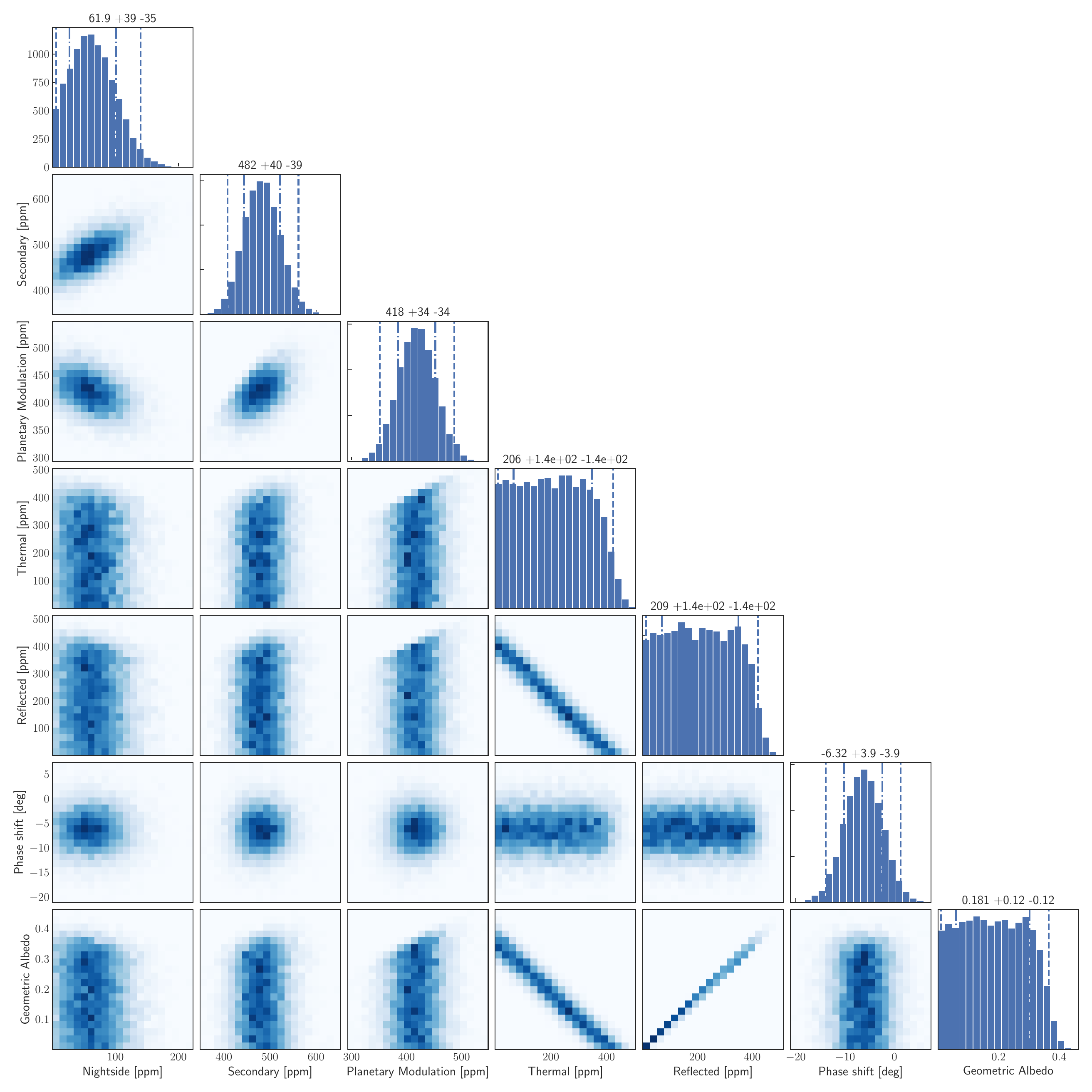}
    \caption{Joint posterior probability distribution of the derived parameters characterizing the WASP-121\,b's \textit{TESS} phase curve. Dashed and dashed-dotted lines indicate 2$\sigma$ and $1\sigma$ credible intervals, respectively.}
    \label{figr:pmar_post_alle}
\end{figure*}

We show in Figure~\ref{figr:pcur_alle} the observed \textit{TESS} light curve and the posterior of our nominal model. We detect a secondary eclipse and a planetary modulation with the amplitudes of \bphasecurveatmosphericTESS{} ppm and \bdepthoccundilTESS{} ppm, respectively. The planetary modulation has a westward shift of \bphasecurveatmosphericshiftTESS{} days, which is consistent with zero. The secondary depth corresponds to a dayside brightness temperature of \tmptdayy{} K in the \textit{TESS} passband. We also measure an ellipsoidal variation with an amplitude of \bphasecurveellipsoidalTESS{} ppm, consistent with zero. The posterior probability distribution of several derived parameters is shown in Figure~\ref{figr:postalle}. We also show in Figure~\ref{figr:pmar_post_alle} the joint posterior probability distribution of the light curve model, indicating that the posterior chain is well converged.

The measured red-optical secondary depth of WASP-121\,b is in general agreement with the previous measurements \citep{Delrez+2016, Evans+2017, KovacsKovacs2019}, as shown in Figure~\ref{figr:spec}. However, based only on the retrieval analysis in \citet{Mikal-Evans+2019}, i.e., \emph{excluding} this \textit{TESS} measurement, we predict a secondary depth of $\sim$300 ppm. This implies that the \textit{TESS} measurement of the secondary depth is above the expectation based on the measurements of the secondary in redder bands and that the dayside emission spectrum of WASP-121\,b is inconsistent with that of a blackbody. 

As for the nightside emission, we find a nightside brightness temperature of \tmptnigh{} K for WASP-121\,b and a westward phase shift of \phasshft{} degree, which is statistically consistent with zero. The measured nightside temperature implies a temperature contrast of \tmptcont{}\%.

\begin{figure*}
    \centering
    \includegraphics[width=0.95\textwidth]{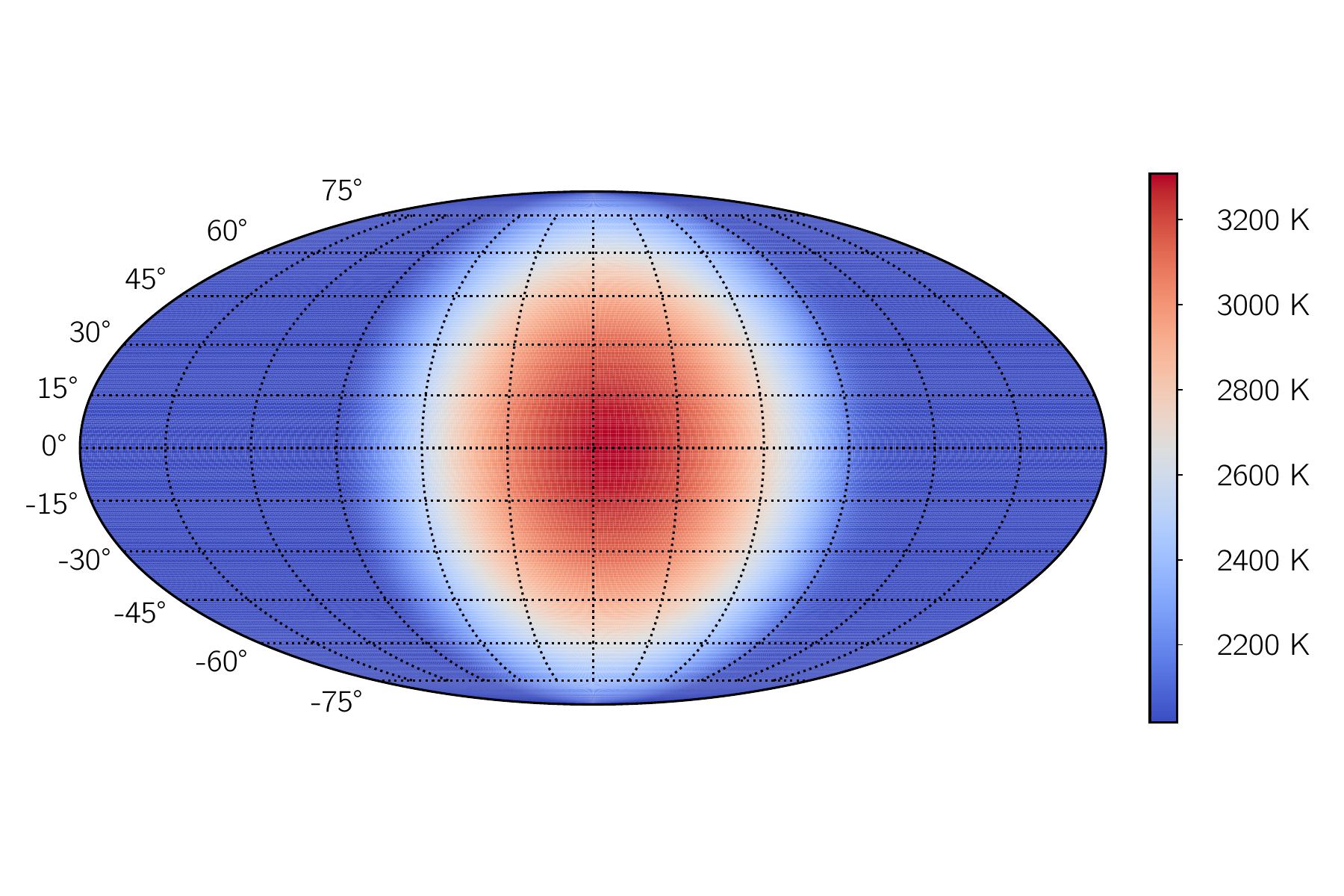}
    \caption{A two dimensional map of the temperature of WASP-121\,b based on the best-fit solution to Equation~\ref{equa:tmpt}.}
    \label{figr:tmpt}
\end{figure*}

Using these inferences, we then solve for the two dimensional temperature distribution on WASP-121\,b as shown in Figure~\ref{figr:tmpt}, which illustrates the lack of the phase shift and reveals the large temperature gradient as a function of longitude. This picture is in contrast with the hot Jupiter HD 189733b, which has a thermal phase shift of $\sim 30$ degrees due to strong equatorial winds \citep{Knutson+2009}.

\subsection{Atmospheric characteristics}

\begin{figure*}
    \centering
    \includegraphics[width=0.95\textwidth]{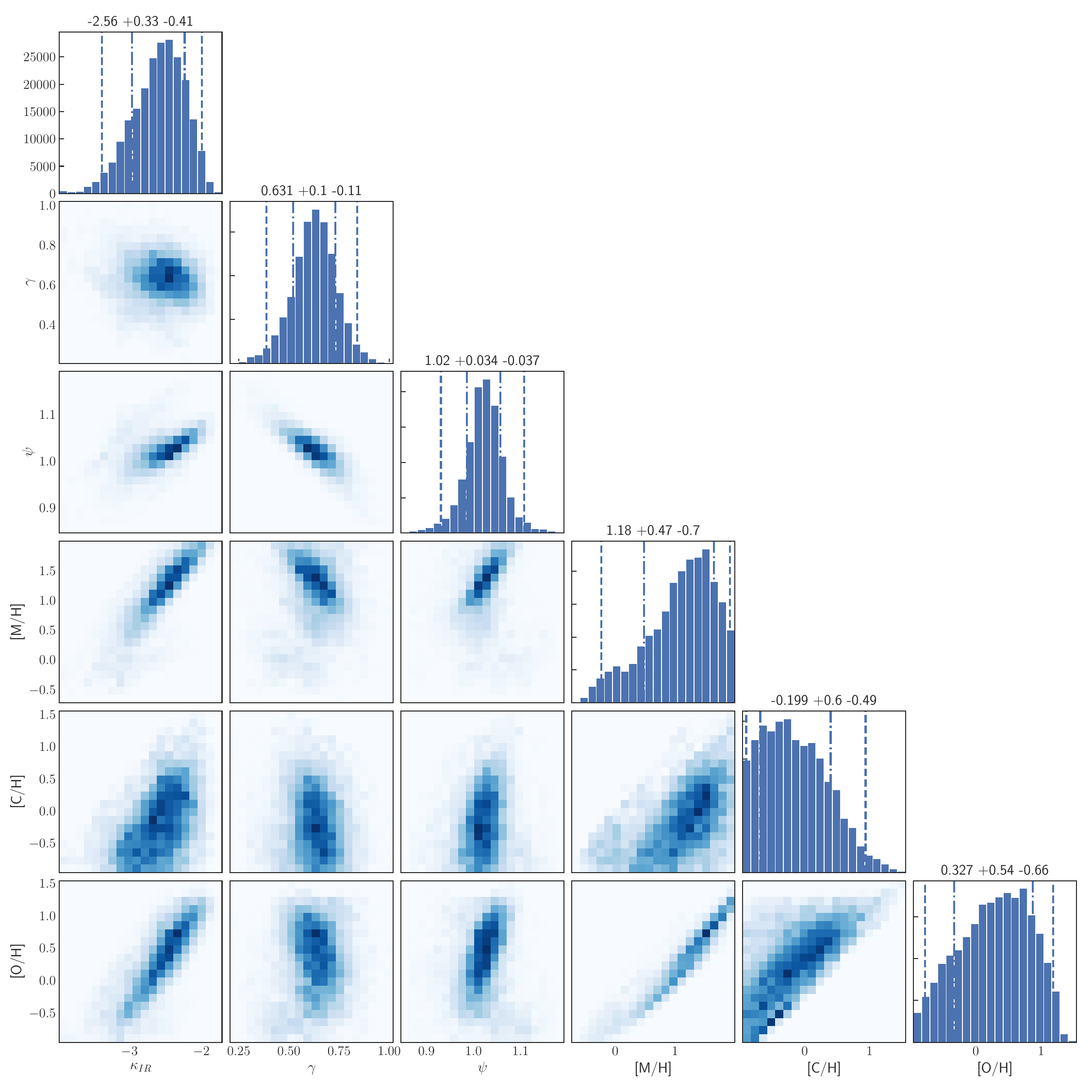}
    \caption{Joint posterior probability distribution of the ATMO parameters for the dayside atmosphere.}
    \label{figr:pmar_post_atmo}
\end{figure*}

\begin{figure*}[!htbp]
    \centering
    \includegraphics[trim=2cm 2cm 0cm 2cm, width=1.05\textwidth]{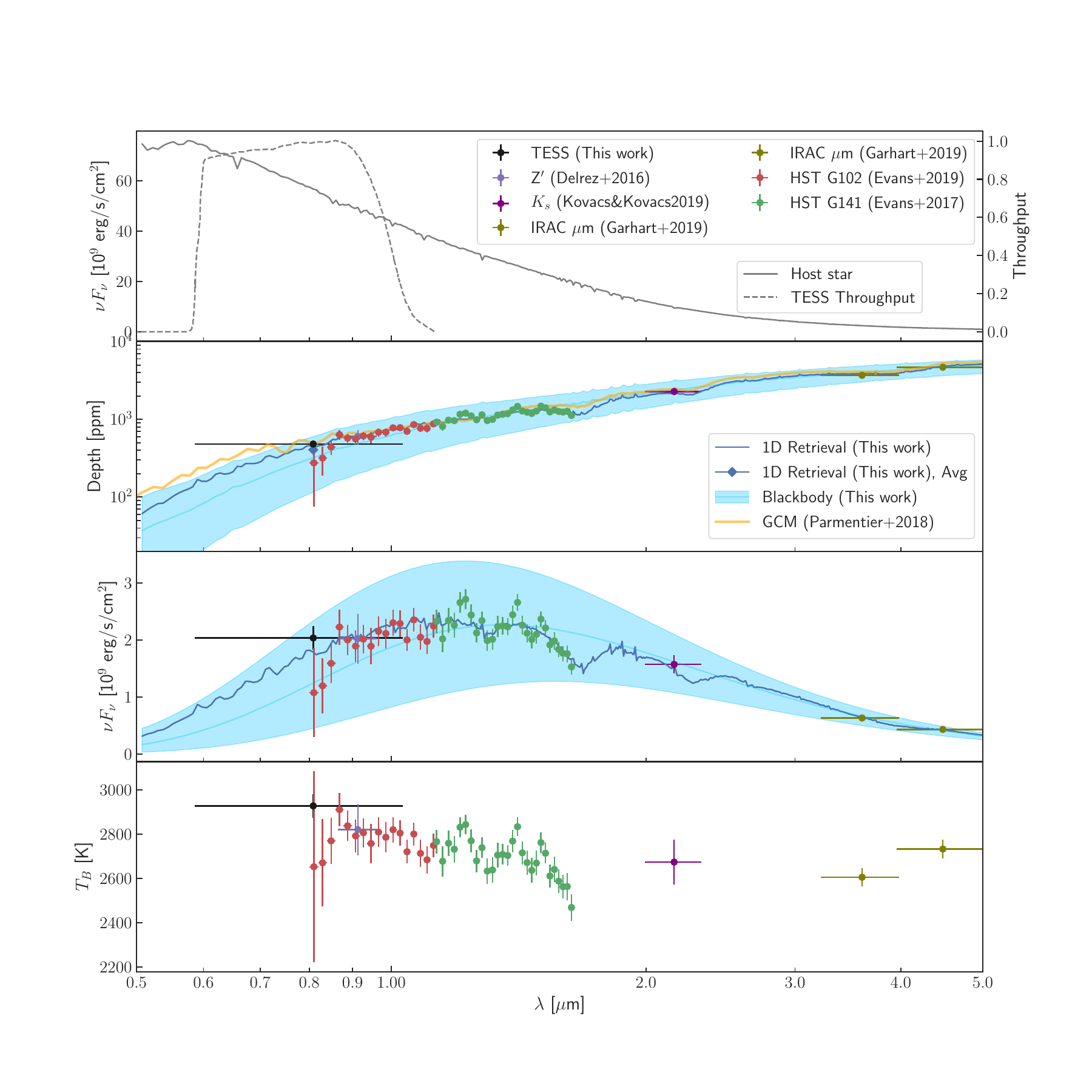}
    \caption{Top: The NextGen model spectrum of the host star, WASP-121  (continuous line) \citep{Hauschildt+1999} and the throughput of \textit{TESS} (dashed line). Second from the top: The measured secondary depths of WASP-121\,b including our atmospheric retrieval model (blue) and the blackbody model (light blue line and uncertainty band). For comparison, we also show the GCM prediction from \citet{Parmentier+2018} (yellow). Second from the bottom: dayside emission spectrum of WASP-121\,b. Bottom: inferred dayside brightness temperature.}
    \label{figr:spec}
\end{figure*}

The posterior probability distribution of our ATMO parameters is given in Figure~\ref{figr:pmar_post_atmo} and Table~\ref{tabl:atmo}. With 44 degrees of freedom, we get a $\chi^2$ per degree of freedom of 1.05, indicating a good fit to the spectral data shown in Figure~\ref{figr:spec}. We note that the discrepancy between the TESS depth and the bluest WFC3 G102 data is likely due to the fact TESS passband extends down to 0.6~$\mu$m and has contribution from species that WFC3 G102 is not sensitive to.

\begin{table}[]
    \centering
    \begin{tabular}{c|c|c}
        Name & Symbol & Value \\
        \hline
        Infrared opacity & $\kappa_{IR}$   & $2.56\substack{+0.33 \\ -0.41}$  \\
        Optical to infrared opacity ratio & $\gamma$        & $0.63\substack{+0.10 \\ -0.11}$  \\
        Irradiation efficiency & $\psi$          & $1.02\substack{+0.03 \\ -0.04}$  \\\relax
        Logarithm of the relative metallicity & [M/H] & $1.18\substack{+0.47 \\ -0.70}$ dex  \\
        Logarithm of the relative C abundance & [C/H] & $-0.20\substack{+0.60 \\ -0.49}$ dex  \\
        Logarithm of the relative O abundance & [O/H] & $1.33\substack{+0.54 \\ -0.66}$ dex
    \end{tabular}
    \caption{Posterior median and 1$\sigma$ uncertainties of ATMO parameters.}
    \label{tabl:atmo}
\end{table}

The atmospheric retrieval indicates that the metallicity of the dayside atmosphere is $\sim$15 times that of the Sun, i.e., $1.18\substack{+0.47 \\ -0.7}$ dex. Furthermore, the abundance of carbon is not well-constrained, especially at the lower end, whereas that of oxygen is constrained to be near the Solar abundance. We note that our high metallicity constraint is not driven by the TESS data point. Indeed, as shown in \citep{Mikal-Evans+2019}, atmospheric retrievals without the TESS data point also lead to high metallicities. The high metallicity is more likely driven by the red end of the HST data that have a steep downward slope reminiscent of the H$^-$ bound-free opacity \citep{Mikal-Evans+2020}.

\begin{figure}
    \centering
    \includegraphics[width=0.45\textwidth]{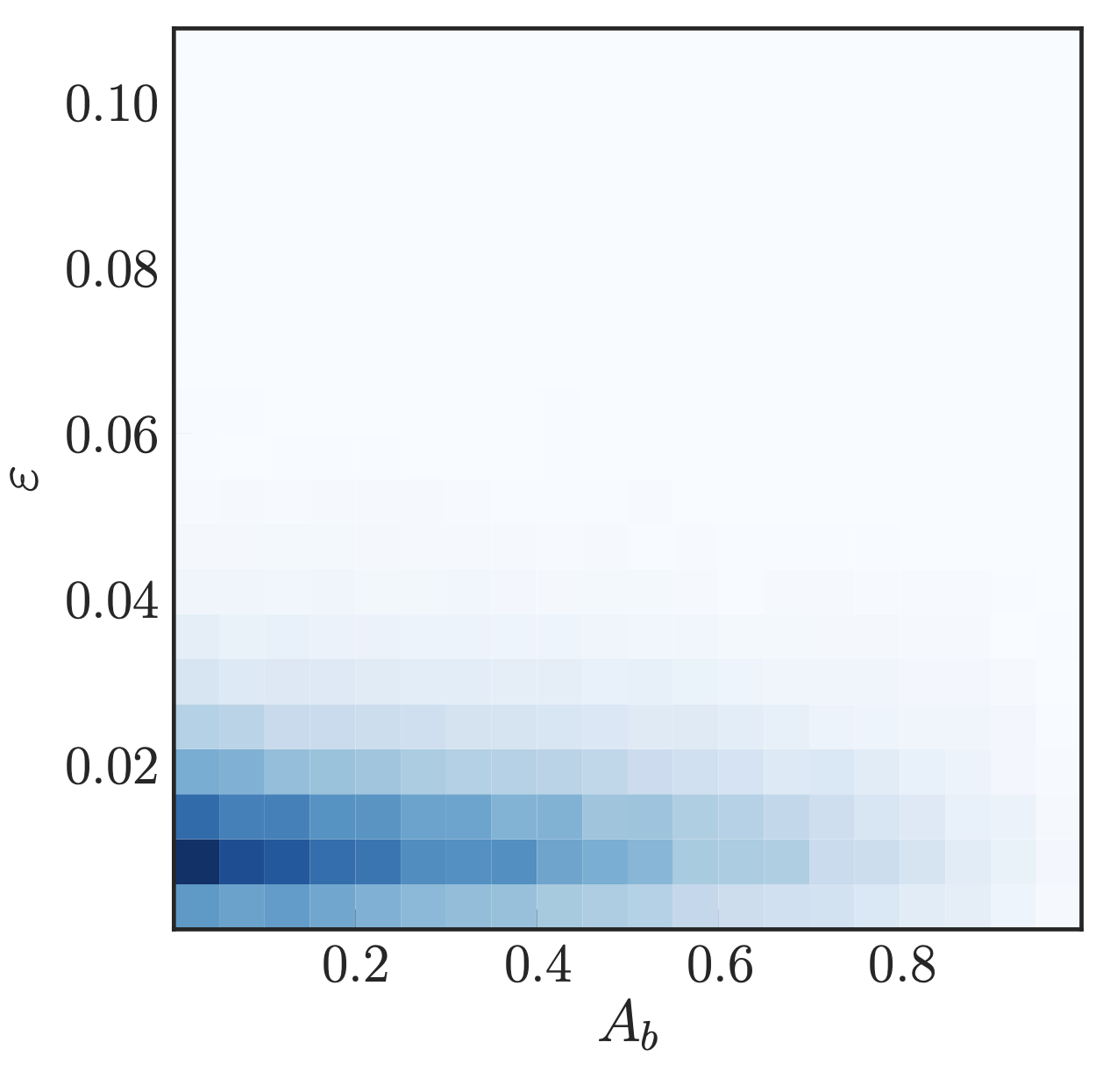}
    \caption{Joint posterior probability distribution of the Bond albedo, $A_{\rm B}$, and heat recirculation efficiency, $\varepsilon$.}
    \label{figr:pmar_post_albbemisepsi}
\end{figure}

The irradiation efficiency, $\psi$, of the dayside is also constrained near unity, indicating that the dayside atmosphere of WASP-121\,b is consistent with being non-reflecting and has inefficient heat recirculation. In order to constrain the latter parameters, we use Equation~\ref{equa:tmptsmaj} and Equation~\ref{equa:tmptdayy} to find the joint posterior probability distribution over the Bond albedo and the heat recirculation efficiency as shown in Figure~\ref{figr:pmar_post_albbemisepsi}.

\begin{figure}
    \centering
    \includegraphics[width=0.45\textwidth]{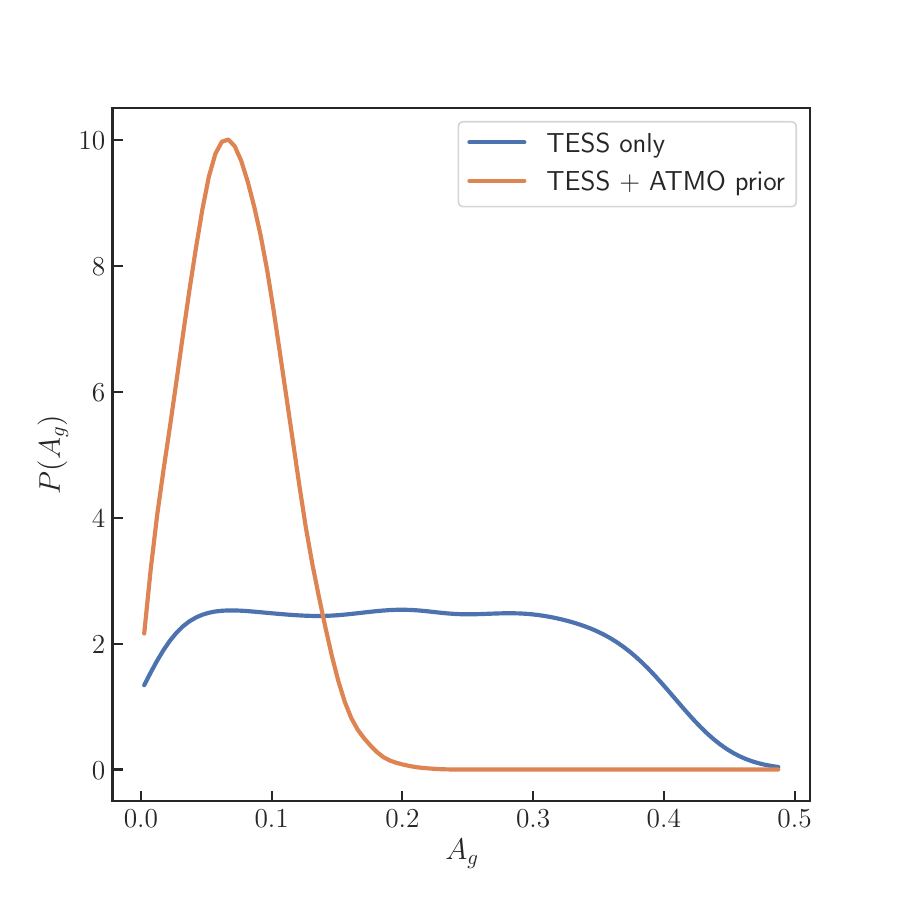}
    \caption{Posterior of the geometric albedo. The blue curve shows the albedo constraint obtained based only on the \textit{TESS} phase curve. The yellow curve shows the posterior informed by the thermal emission prediction from ATMO.}
    \label{figr:pdfn_albg}
\end{figure}

Using the \textit{TESS}-informed, predicted emission spectrum of WASP-121\,b as shown in Figure~\ref{figr:spec}, we then solve for the geometric albedo in the \textit{TESS} passband. An estimate for WASP-121\,b's geometric albedo in the $z^\prime$ band was presented in \citep{Mallonn2019} as $A_{\rm g} = 0.16 \pm 0.11$. We find a geometric albedo of $0.07\substack{+0.037 \\ -0.040}$ in the \textit{TESS} band. This reveals mild (i.e., $\sim 2\sigma$) evidence that the atmosphere of WASP-121\,b has non-negligible reflectivity in the \textit{TESS} passband. Figure~\ref{figr:pdfn_albg} compares the geometric albedo inferred based on only the \textit{TESS} data and that inferred incorporating the thermal emission constraint from the atmospheric retrieval.

\begin{figure*}[!htbp]
    \centering
    \includegraphics[width=1.05\textwidth]{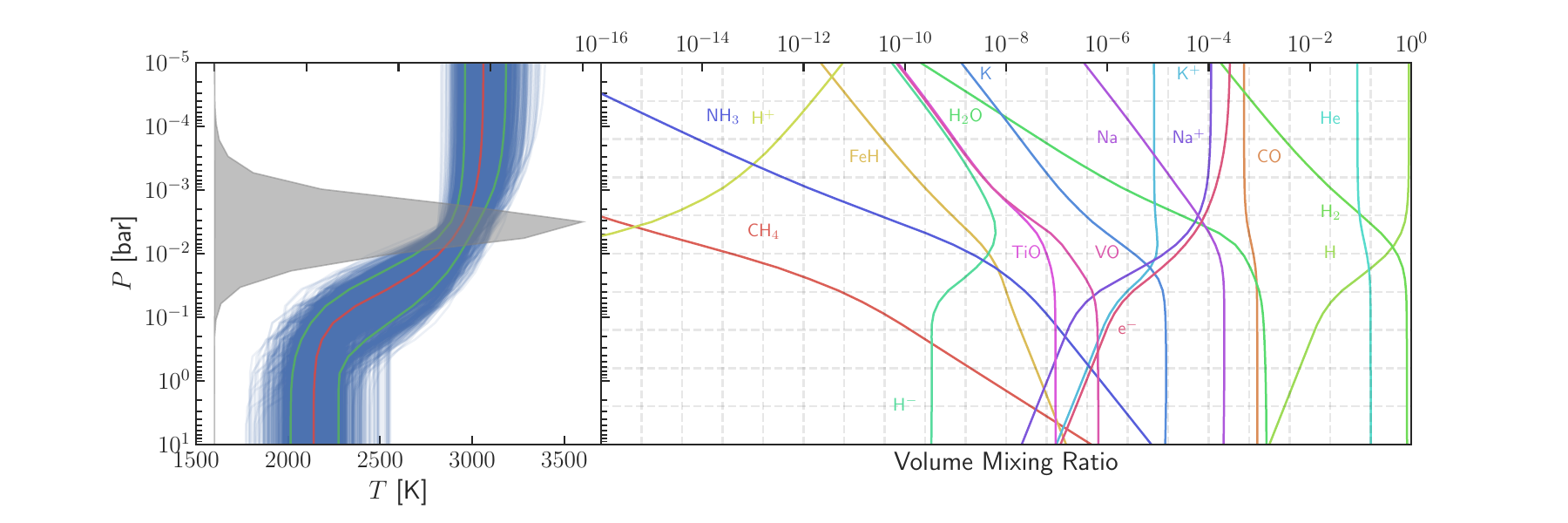}
    \caption{Left: Pressure-temperature profile of the dayside atmosphere of WASP-121\,b. Blue lines are fair draws from the posterior, green lines are the 16th and 84th percentiles and the red line indicates the median. The superimposed filled grey curve shows the contribution function of \textit{TESS}, revealing the altitude range for which the \textit{TESS} data is informative. Right: Volume mixing ratios of various chemical species in the dayside atmosphere of WASP-121\,b inferred based on the atmospheric retrieval.}
    \label{figr:ptem}
\end{figure*}

The radiative transfer modeling also allows us to constrain the pressure-dependent temperature profile of the atmosphere, abundances of chemical species and the contribution function of \textit{TESS}. In Figure~\ref{figr:ptem}, we present the pressure-temperature profile. In \citep{Mikal-Evans+2019} the atmospheric temperature was found to vary from 2500 K up to 2800 K from 30 mbar down to 5 mbar pressure. We find that the top of WASP-121\,b's atmosphere is hotter than previously thought and rises well above 3000 K.

The informativeness about the atmosphere of a given passband is encapsulated in a \emph{contribution function}. This function indicates how much flux a particular wavelength interval contributes to each horizontal slice of the atmosphere.

\begin{figure}[!htbp]
    \centering
    \includegraphics[width=0.45\textwidth]{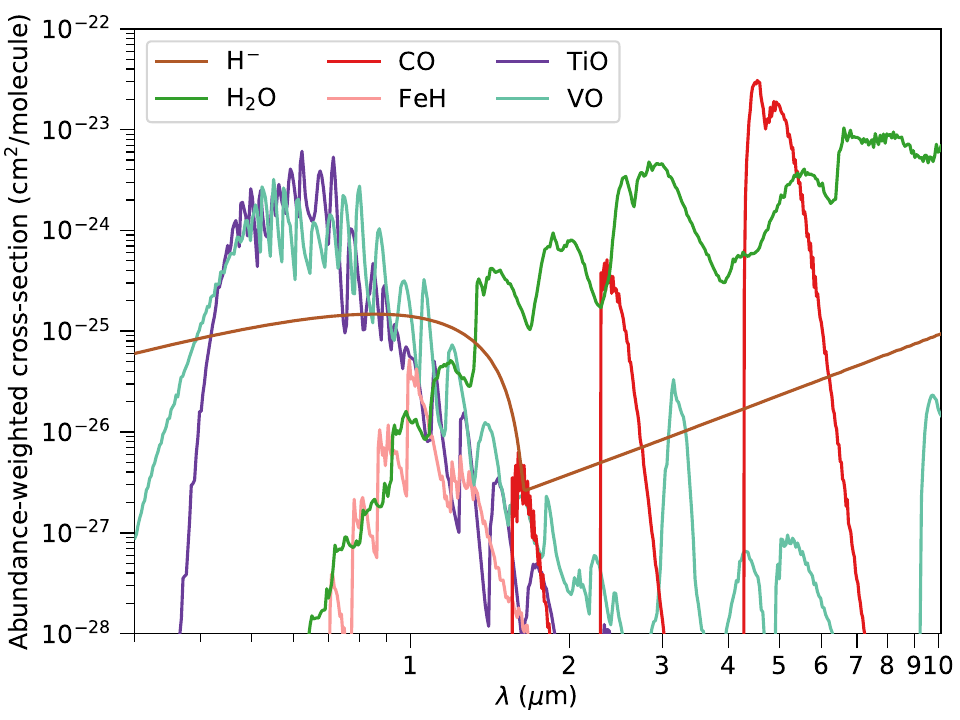}
    \caption{Cross sections of relevant species, weighted by their abundances at a pressure of 10 mbar. H$^-$, TiO and VO are the most important potential contributors to the opacity in the \textit{TESS} passband.}
    \label{figr:csec}
\end{figure}

We show in Figure~\ref{figr:csec} the abundance-weighted cross section of important species. We find that H$^-$, TiO and VO have the highest contribution to the opacity in the \textit{TESS} passband at chemical equilibrium. The arrival of the \textit{TESS} data raises their abundance-weighted cross section by roughly an order of magnitude and results in a dayside atmosphere with a much more pronounced temperature inversion.

\section{Discussion}
\label{sect:disc}

The \textit{TESS} phase curve of WASP-121\,b provides further evidence for the temperature inversion of its dayside atmosphere. The measured secondary depth also implies that the \textit{TESS} secondary is deeper than expected based on previous measurements at longer wavelengths. There may be two components of this excess.

First, the planet may be reflecting some light in the \textit{TESS} passband. Given that \textit{TESS} passband covers shorter wavelengths, where the exoplanet-to-star contrast of thermal emission is lower, it is reasonable that this excess is due to reflected light from the exoplanet. In general, condensates such as H$_2$O and NH$_3$ on relatively cool (e.g., $\lesssim 300$ K) planets or condensates (e.g., silicates, Al$_2$O$_3$, CaTiO$_3$) on hotter planets (e.g., $\gtrsim 1500$ K) can make a planet reflective and result in a large albedo. However, ultra-hot Jupiters such as WASP-121\,b are not expected to have a high reflectivity \citep{Bell2017, Shporer+2019} due to Mie scattering by clouds, because condensates required for this type of reflection would not form due to the high temperatures. Furthermore, cloud-free Rayleigh scattering is only effective at short wavelengths and would not be enough to increase the albedo to the inferred value in the \textit{TESS} passband. However, despite the large substellar temperature of WASP-121\,b, reflection can still happen from condensates on the western terminator \citep{Parmentier+2015}.

Secondly, the excess may be due to additional sources of opacity in the dayside atmosphere of WASP-121\,b with emission lines in the \textit{TESS} passband, causing the planet to be heated at high altitudes. This would be consistent with earlier findings that there is a temperature inversion in the dayside atmosphere. This is also supported by our atmospheric retrieval, which predicts that the inversion in the pressure-temperature profile is more pronounced and that the atmosphere is hotter at high altitudes compared to \citet{Mikal-Evans+2019}. The potential temperature inversion is also consistent with the pressure temperature profile in \citet{Parmentier+2018}. Potential species that can cause this additional absorption are continuum opacity due to $H^-$, metals such as Fe gas \citep{Lothringer+2018}, or band opacity due to TiO or VO. Although a minor effect, the internal heat of the planet might play a role in increasing the temperature as well. Finally, the presence of a spatially nonhomogeneous thermal structure on the dayside could be leading to an increase of the thermal emission at short wavelengths \citep{Taylor+2020}.

The allowed range of heat recirculation can be explained as heat can be transported from the dayside to the nightside when molecular hydrogen, H$_2$, thermally dissociates into atomic hydrogen, H, recombines on the nightside into H$_2$, and releases internal energy. This process can make heat recirculation more efficient than that based on zonal jets only \citep{BellCowan2018, KomacekTan2018}. A known trend in ultra-hot Jupiters is the increase of the day-night temperature contrast with the increase of dayside temperatures. This is due to the radiative time scale being too short in ultra-hot Jupiters \citep{Komacek+2016}. The expected resulting dayside-to-nightside fractional temperature contrast for WASP-121\,b is $\sim$30\%, which is consistent with what we observe. We used \textsf{Bell\_EBM}\footnote{https://github.com/taylorbell57/Bell\_EBM} \citep{BellCowan2018} to model the energy balance of WASP-121\,b including the effect of hydrogen dissociation and recombination. In this model, the predicted planetary modulation amplitude and phase shift are $\sim$600 ppm and $\sim$40 degrees eastward, respectively. The latter prediction is inconsistent with the observed phase shift and requires a non-negligible contribution from reflected light in the \textit{TESS} passband, as also suggested by the fact that the secondary eclipse depth measured by \textit{TESS} is larger than that predicted by ATMO based on multiband data. Our results indicate that WASP-121\,b indeed has a phase shift consistent with zero. This could arise from competing phase shifts induced by the thermal and reflection processes mentioned above, which may be canceling each other.

We further compare our phase curve with the Global Circulation Models (GCMs) extensively described in \citet{Parmentier+2018}. The SPARC/MITgcm solves the primitive equation of hydrodynamics on a cube-sphere grid and the non-grey radiative transfer equations \citep{Showman+2009}, assuming chemical equilibrium (including the thermal dissociation of molecules) and using the opacities from \citet{Freedman+2014}. The resulting outgoing flux is integrated over the \textit{TESS} passband and compared to the observation in Figure~\ref{figr:pcur_alle}. Despite its complexity the model presented here has several shortcomings: it neglects, among others, clouds, magneto-hydrodynamics effects and day/night latent heat transport through H$_2$ dissociation. Nonetheless, without adjusting any parameters the model predicts a reasonable phase curve amplitude in the \textit{TESS} passband. In more detail, the model predicts a larger than observed phase curve offset and smaller than observed dayside flux. This could be the sign that magnetic drag is significantly slowing the winds of this planet compared to the simulation, as was proposed for WASP-18b \citep{Arcangeli+2019} and WASP-103b \citep{Kreidberg+2018}. Additionally, the model underpredicts the nightside flux in the \textit{TESS} passband, pointing towards the presence of an additional source of day-night heat transfer, most likely the transfer of energy through the latent heat of the dissociation/recombination of H$_2$.

The dayside emission spectrum predicted by the GCM also provides a good match 
to both the WFC3/G102 and the \textit{TESS} data as shown in Figure~\ref{figr:spec}. The red end of the WFC3/G121 
dataset is poorly fitted by the GCM, which was already discussed in \citet{Parmentier+2018} and \citet{Mikal-Evans+2019}. As shown by the retrievals, the WFC3/G121 datasets points toward a non-solar composition. An exploration of the parameter space with the GCM by varying the atmospheric metallicity and elemental abundance ratio would probably allow us to find a better match to the data, but is out of the scope of this paper.

The ellipsoidal variation was not measured to be statistically significant in the WASP-121 system. This amplitude of the ellipsoidal variation likely gets contributions from the tidal distortion of both the host star and WASP-121\,b. The cross-sectional area for WASP-121\,b should vary by $\sim$10\% over the course of its orbit, implying that a large fraction of the ellipsoidal variation can be due to the tidal distortions of the planet \citep[e.g., ][]{Kreidberg+2018}.

Future work on this system can potentially characterize the atmospheric temperature distribution to a greater detail (i.e., perform eclipse mapping) using GCMs. Magnetohydrodynamical (MHD) effects can also be investigated \citep{RogersShowman2014}, as WASP-121\,b's high substellar temperature causes a large amount of ionization as shown in the right panel of Figure~\ref{figr:ptem}.

\section{Conclusion}
\label{sect:conc}

In this work, we present the first analysis of the full phase curve of the ultra-hot Jupiter WASP-121\,b as measured by \textit{TESS}. We characterize the phase curve and the dayside atmospheric properties of WASP-121\,b and find further evidence for its temperature inversion. We find that \textit{TESS} observes excess emission from the dayside of WASP-121\,b in the red-optical passband, which could be attributed to higher-than-expected reflected light or additional optically-thick absorbers in the red-optical band. We also measure the night-side emission, the temperature contrast, and the phase shift between the substellar point and the hot spot, finding that the heat transport on WASP-121\,b must be inefficient.

\section*{Acknowledgments}
Funding for the \textit{TESS} mission is provided by NASA's Science Mission directorate. 
This paper includes data collected by the \textit{TESS} mission, which are publicly available from the Mikulski Archive for Space Telescopes (MAST).

Resources supporting this work were provided by the NASA High-End Computing (HEC) Program through the NASA Advanced Supercomputing (NAS) Division at Ames Research Center for the production of the SPOC data products.

TD acknowledges support from MIT's Kavli Institute as a Kavli postdoctoral fellow.
MNG acknowledges support from MIT's Kavli Institute as a Torres postdoctoral fellow.
Work by JNW was partly supported by the Heising-Simons Foundation.
DC acknowledges support from the John Templeton Foundation. The opinions expressed in this publication are those of the authors and do not necessarily reflect the views of the John Templeton Foundation.

We thank the anonymous reviewer for the useful comments and suggestions during the revision of the manuscript.

{
\textit{Facilities}:
{\textit{TESS}}

\textit{Software}:
\texttt{python} \citep{Rossum1995},
\texttt{matplotlib} \citep{Hunter2007},
\texttt{seaborn}
(\url{https://seaborn.pydata.org/index.html}),
\texttt{numpy} \citep{vanderWalt2011},
\texttt{scipy} \citep{Jones2001}, 
\texttt{allesfitter} \citep{GuentherDaylan2019a, GuentherDaylan2019b},
\texttt{ellc} \citep{Maxted2016},
\texttt{emcee} \citep{Foreman-Mackey2013},
\texttt{celerite} \citep{Foreman-Mackey2017},
\texttt{corner} \citep{Foreman-Mackey2016}.
\texttt{dynesty}\citep{Speagle2019},
}

\bibliography{refr}

\end{document}